\documentclass[10pt,journal,compsoc]{IEEEtran}

\usepackage[american]{babel}
\usepackage{comment}
\usepackage{graphicx}

\usepackage{paralist}

\usepackage[math]{blindtext}

\usepackage{amsthm}
\usepackage{amsmath}
\usepackage{algorithm}
\usepackage[noend]{algpseudocode}

\makeatletter
\def\BState{\State\hskip-\ALG@thistlm}
\makeatother

\usepackage[
bookmarks=false,
breaklinks=true,
colorlinks=true,
linkcolor=black,
citecolor=black,
urlcolor=black,
pdfpagelayout=SinglePage
]{hyperref}
\usepackage[all]{hypcap}

\usepackage[capitalise,nameinlink]{cleveref}
\crefname{section}{Sect.}{Sect.}
\Crefname{section}{Section}{Sections}
\crefname{figure}{Fig.}{Fig.}
\Crefname{figure}{Figure}{Figures}

\usepackage{xspace}

\newcommand{\fullversion}[1]{}
\newcommand{\extabs}[1]{#1}
\newcommand{\swallow}[1]{}

\DeclareFontFamily{U}{MnSymbolC}{}
\DeclareSymbolFont{MnSyC}{U}{MnSymbolC}{m}{n}
\DeclareFontShape{U}{MnSymbolC}{m}{n}{
    <-6>  MnSymbolC5
   <6-7>  MnSymbolC6
   <7-8>  MnSymbolC7
   <8-9>  MnSymbolC8
   <9-10> MnSymbolC9
  <10-12> MnSymbolC10
  <12->   MnSymbolC12%
}{}
\DeclareMathSymbol{\powerset}{\mathord}{MnSyC}{180}

\makeatletter
\g@addto@macro{\UrlBreaks}{\UrlOrds}
\makeatother

\ifCLASSOPTIONcompsoc
  \usepackage[nocompress]{cite}
\else
  \usepackage{cite}
\fi

\usepackage{array}

\ifCLASSOPTIONcompsoc
  \usepackage[caption=false,font=footnotesize,labelfont=sf,textfont=sf]{subfig}
\else
  \usepackage[caption=false,font=footnotesize]{subfig}
\fi

\usepackage{fixltx2e}
\usepackage{dblfloatfix}

\newtheorem{theorem}{Theorem}
\newtheorem{lemma}{Lemma}
\newtheorem{corollary}{Corollary}

\begin{document}

\input glyphtounicode.tex
\pdfgentounicode=1

\title{Live Phylogeny with Polytomies: Finding the Most Compact Parsimonious Trees}

%
\author{Dimitris~Papamichail, Angela~Huang, Edward~Kennedy, Jan-Lucas~Ott, Andrew~Miller, Georgios~Papamichail%
\IEEEcompsocitemizethanks{\IEEEcompsocthanksitem D. Papamichail, A. Huang, E. Kennedy, J. Ott and A. Miller are with the Department
of Computer Science, The College of New Jersey, Ewing,
NJ, 08628.\protect\\
E-mail: papamicd@tcnj.edu
\IEEEcompsocthanksitem G. Papamichail is with the Department of Computer Science, New York College, Athens, Greece.}%
}
            

\IEEEtitleabstractindextext{%
\begin{abstract}
Construction of phylogenetic trees has traditionally focused on binary trees where 
all species appear on leaves, a problem for which numerous efficient solutions
have been developed. Certain application domains though, such as viral evolution 
and transmission, paleontology, linguistics, and phylogenetic stemmatics, often require 
phylogeny inference that involves placing input species on ancestral tree nodes (live
phylogeny), and polytomies.
These requirements, despite 
their prevalence, lead to computationally harder algorithmic solutions and
have been sparsely examined in the literature to date.
In this article we prove some unique properties of most parsimonious live phylogenetic trees with polytomies, 
and describe novel algorithms 
to find the such trees without 
resorting to exhaustive enumeration of all possible tree topologies.
\end{abstract}

\begin{IEEEkeywords}
Phylogenetics, Maximum Parsimony, Live Phylogeny, Polytomies.
\end{IEEEkeywords}}

\maketitle

\IEEEdisplaynontitleabstractindextext
\IEEEpeerreviewmaketitle

\section{Introduction}\label{sec:intro}

Phylogeny is the evolutionary history of a set of species whose relationships are often represented by a tree.
Phylogenetic trees can be rooted or unrooted, and their edges are labelled with lengths that correspond to evolutionary
distances between species. 

\fullversion{
There are several prominent tree construction methods for evolutionary inference. {\it 
Distance Matrix} methods involve the generation of dissimilarity matrices based on 
well-defined distances, and the application of clustering algorithms such as 
Neighbor-Joining \cite{Sai87}. {\it Maximum Parsimony} is a 
method that uses characters instead of distances, associates a cost with each 
character mutation (event), and aims to build a tree with the smallest possible cost. {\it Maximum 
Likelihood} and {\it Bayesian} methods involve assigning probabilities to events and probability 
distributions to possible tree configurations, aiming to create trees whose observed 
phylogeny has the highest likelihood. In this study we will use the Maximum Parsimony 
method with unordered character states, a well studied, simple, and intuitive 
criterion.
}

\extabs{
{\it Maximum Parsimony} is a
method that uses {\it characters}, associates a cost with each
character mutation ({\it event}), and aims to build a tree with the smallest possible cost.
In recent years, statistical methods \cite{Ron01,Fel81} have supplanted maximum parsimony 
approaches for constructing phylogenies in certain domains. However, maximum parsimony
remains an effective and widely-used method to predict correct
viral phylogenies based on genomic data \cite{Hil92,Zhu98,Bus99}, for morphological characters \cite{Wil03}, to build supertrees \cite{Sal02}, and to perform fast heuristic tree searches \cite{Swo09}.
}

This article focuses on phylogenies where ancestors can be present among the input species, a concept termed {\it live phylogeny} by Telles et al. in \cite{Tel13}. 
Existing phylogenetic methods have primarily focused on fully bifurcating trees where all extant species are placed on the 
leaves of the tree. However, in domains such as virology, paleontology, linguistics, and phylogenetic stemmatics, it is often the case that internal ancestor nodes can be either hypothetical or input species. The ability
to identify known common ancestors using molecular data has been successfully demonstrated with the Ebolavirus and 
Marburgvirus genera \cite{Car13}. Patterns of evolution of HIV within patients have been shown to detect emergence of 
specific strains \cite{Bue09}, using {\it serial evolution networks}, which resemble trees with extant ancestor nodes. 
In the area of 
paleontology, ancestors of species may be known and well characterized, prompting the need for phylogenetic 
reconstruction methods that account for labeled internal nodes. Notably, the fossil record is incomplete, and it does not 
provide a high guarantee of recording the common ancestor of species \cite{Foo96}. However, there are certain species where
the fossil record has been extensively studied and extinct common ancestors are highly known, such as the case for 
graptolites (e.g. \cite{Mie01,Urb98}). Existing efforts to build trees which incorporate 
known ancestors, such as the paleotree package \cite{Bap13}, can greatly benefit from the
algorithmic methods presented in this paper.

\extabs{
Besides allowing for input species to appear on internal nodes, it is also important in 
certain domains, such as viral transmission and phylogenetic stemmatics, to account for 
polytomies, utilizing multifurcating trees instead of 
strictly bifurcating ones. 
For example, in a study of phylogenies that were reconstructed from 38 different RNA 
viruses, all phylogenies contained a number of polytomies. Forcing the polytomy to a 
bifurcating structure due to limitations in the implemented algorithm added a source 
of uncertainty to the phylogenetic reconstructions \cite {Wal13}.
Some previous work in polytomies focused on constant
time heuristic improvements \cite{Gol02}; our work instead focuses on native methods 
for identifying the most parsimonious tree allowing for polytomies.
Lack of work in this area may be a result of the additional complexity polytomies add 
to an already hard computational problem \cite{Gra82,Day83,Day86,Car14}.
}

With this work, we aim to explore the construction of maximum parsimony trees that 
allow for polytomies and internal species nodes. Such trees 
have been named {\it X-trees} by Steel et al. and certain of their properties have been examined in \cite{Sem03}.
Mapping species to internal nodes reduces tree size, as do edge contractions among internal nodes,
which introduce (or increase the degree of) polytomies. As such, minimization of the 
number of nodes in a tree with $n$ species becomes now an additional parsimony criterion 
to the number of events along the edges, as we aim to create the 
most compact parsimonious trees.

The rest of the paper is structured as follows: In section \ref{sec:defs} we provide 
terminology for most terms encountered in this paper. Section \ref{sec:enum} examines 
the number of phylogenetic trees with $n$ species that make up our search space, and 
compares its magnitude to the number of cubic trees with $n$ species, which is 
explored in traditional phylogenetic algorithms. 
\extabs{In section \ref{sec:small_pars} we 
describe Hartigan's algorithm, which solves the small parsimony 
problem with polytomies, and adapt it from rooted to unrooted trees.}
\fullversion{In section \ref{sec:small_pars} we 
describe the algorithms of Fitch and Hartigan, which solve the small parsimony 
problem, and adapt the latter from rooted to unrooted trees.}
In section \ref{sec:compact} we 
describe an algorithm to find the most compact parsimonious tree using edge 
contractions. We present results that demonstrate the efficiency of the contraction algorithm
in section \ref{sec:results} and conclude with observations and discussion in 
section \ref{sec:conclusion}.

\section{Definitions}\label{sec:defs}

A rooted tree where all nodes have a maximum degree of $3$ is called a {\it binary} or {\it bifurcating} tree.
If all internal nodes except for the root have a degree of 3 (one parent and two children) then the rooted tree
is called a {\it full binary tree}. An unrooted tree where all nodes have either a degree of 1 (leaves) or 3 (internal
nodes) we will call a {\it cubic tree}, following the terminology of \cite{Exo96}. A tree whose nodes
can have degrees $> 3$ is called {\it multifurcating}. Nodes in a tree can be {\it labelled}, i.e. assigned
values. A {\it labelled-leaf tree} has values assigned to all of its leaves. In our study we will define a 
{\it mixed-labelled tree} (or {\it mixed tree}) to be a tree where all leaves are labelled, and internal nodes may be labelled.

\fullversion{
\subsection{Node labels, characters and states}\label{subsec:char}
}

The following definitions follow to a large extent the terminology in \cite{Har73}.
Let $S$ be a set of $n$ objects $\{S_1, S_2, \dots, S_n\}$.
We will refer to these objects as {\it species}\fullversion{, independent of whether they
represent organisms, documents, or any other kind of objects whose evolutionary
relationships we are attempting to infer}.
Each species has a set of $m$ ordered features $C$, called {\it characters}. 
\fullversion{In organisms, such
characters can be observed heritable traits, often morphological or genomic. In the
comparison of documents, characters often are words or sections of manuscripts.
}
Each character can take a constant number of values, called {\it states}.
\fullversion{For example, a DNA character can have four states, $\{A, C, G, T\}$, where an amino acid character can have 20.
}

Each species $S_i, 1 \le i \le n$ is a fixed tuple of $m$-character states $(C(i)_1, C(i)_2, $
$\dots, C(i)_m)$.
\swallow{For our study each character $C_j, 1 \le j \le m$ can have a different fixed number of states.} 
Character states are unordered (Fitch parsimony).
Species can be assigned to nodes in a tree, which are then considered labelled. As such, every labelled (species) node in a tree
will have an m-tuple of character states associated with it, which will be the {\it value} $V$ of the node.
Each labelled node $v$ will also have a {\it root set} $VV$ associated with it,
which is an $m$-tuple of character state singleton sets.
For example,
a labelled node $v_i$ corresponding to species $S_j$ will have a value $V(i) = (A, B, \dots, Z)$ and a
root set $VV(i) = (\{A\}, \{B\}, \dots, \{Z\})$, where
$C(j)_1 = A, C(j)_2 = B, \dots, C(j)_m = Z$. Unlabelled nodes in the
tree will also have root sets $VV$, whose state sets can contain more than one state. If an unlabelled
node $u$ is assigned a single state for each character, then the node is called {\it fitted} and the assignment is called a {\it node fit} $f$, with 
$f \in VV(u)_1 \times VV(u)_2 \times \dots \times VV(u)_m$. 
A {\it tree fit} is an assignment of node fits to all unlabelled nodes in the tree.
\fullversion{
When representing the root set $VV$ of a node, we will use the shorthand notation $A/BCD/EF/\dots/YZ$ instead of the set notation 
$(\{A\}, \{B, C, D\}, \{E, F\}, \dots, \{Y, Z\})$.
}

A {\it mutation} or {\it event} is a change between states of a character. A single mutation will carry a unit {\it cost}.
\swallow{
A distance $d(u, v)$ between two nodes $u$ and $v$ in a tree is defined as the sum of all
mutations among all characters in the values of the nodes. The distance between two adjacent nodes
$u$ and $v$ is assigned to the edge $(u, v)$.
}
Let $nei_i(x,y):X_i\times X_i \to \{0,1\}$, where $X_i$ is the powerset of the states of character $C_i$, 
be a function such that
\[
     nei(x, y)=\left\{\begin{array}{ll}
     0 \text{ if } x \cap y \ne \emptyset \\
     1 \text{ otherwise}
     \end{array}\right.
\]
The {\it minimum distance} $md(u, v)$ between two adjacent nodes $u$ and $v$ is defined as 
\[ md(u, v) = \sum_{1 \le i \le m}{nei_i(VV(u)_i,VV(v)_i)} \]
The {\it potential cost} of an edge $(u, v)$, \swallow{denoted by $pc(u, v)$ }is the number of mutations between a pair of fits of $u$ and $v$.
The {\it cost} of an edge $(u, v)$ is the number of mutations between the values of $u$ and $v$.
The {\it minimum cost} or {\it min-cost} of an edge $(u, v)$\swallow{, denoted by $mc(u, v)$,} is defined as 
the minimum number of mutations between all pairs of fits between $u$ and $v$, and is equal to $md(u, v)$.
The {\it cost} of a tree fit is the sum of costs along the tree's edges.
The {\it most parsimonious cost (MP-cost)} of a tree is the minimum sum of potential costs along all of its edges for any tree fit. An MP-cost tree fit is called a {\it best fit}\fullversion{, and the corresponding tree an {\it MP-tree}}.

\section{Enumerating mixed trees}\label{sec:enum}

According to Flight \cite{Fli90}, there are $\displaystyle\sum_{m=0}^{n-2}T(n, m)$ unrooted mixed 
labelled trees, where all leaf nodes are labelled, and internal nodes may be labelled,
where $T(n, m)$ is the number of unique trees with n labelled nodes and m unlabelled nodes. 
Observably, there are four different ways to construct a tree with $n$ labelled species 
from a tree with $n-1$ labelled species, allowing polytomies:

\begin{enumerate}
\item Insert an unlabelled node into any of the $n+m-3$ edges of any of the $T(n-1, m-1)$ 
trees and have the $nth$ labelled node descend from it.
\item Insert the labelled node directly into any of the $n+m-2$ edges of any of the $T(n-1, m)$ trees. 
\item Make the labelled node the child of any of the $n+m-1$ available nodes belonging to any of the $T(n-1, m)$. 
\item Label any of the $m+1$ unlabelled nodes in any of the $T(n-1, m+1)$ trees. 
\end{enumerate}

This leads to the following recurrence: 
   \[
     \textrm{T}(n, m)=\left\{\begin{array}{ll}
     \quad a\cdot T(n-1,m-1) \\
     +\;b\cdot T(n-1, m)  \\
     +\;c\cdot T(n-1, m+1)
     \end{array}\right.
     \]
where $a=m+n-3$ if $m>0$ or $a=0$ otherwise, $b=2n+2m-3$, and $c=m+1$ if $n>m+2$ or $c=0$ otherwise.
The base case of this recurrence is: $T(1, 0)=1$ and $T(1, i)=0$ $ \forall_i, i>0$. 
Utilizing sequence A005263 from
N.J.A. Sloane's Online Encyclopedia of Integer Sequences \cite{Oei10} we identified the following 
closed form formula as an approximation for the number
of trees as a function of the labelled nodes $n$: 
$\frac{\displaystyle n^{n-2}}{\displaystyle \sqrt{2}e^{\frac{n}{2}}\left ( 2-e^{\frac{1}{2}} \right )^{n-\frac{3}{2}}}$

\begin{figure}
\begin{center}
\includegraphics[width=9cm]{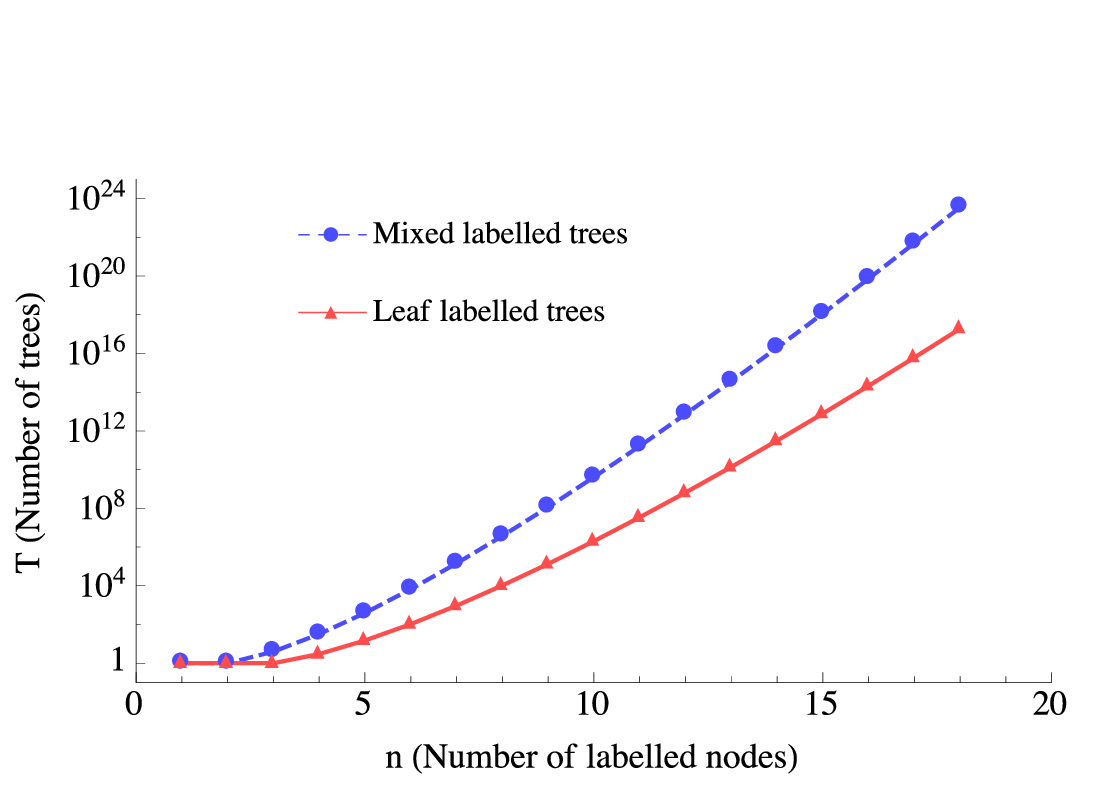}
\end{center}
\caption{Comparison of growth rates of mixed and leaf labelled trees.}
\label{fig:plot_trees}
\end{figure}

\swallow{From observing the growth rate of this function, it is clear that an exhaustive search on all 
possible trees is only practical for small values of $n$. Already, t}The exhaustive search method 
on cubic trees with labelled leaves and unlabelled internal nodes is computationally impractical 
for any but the smallest input sets \cite{Swo09}. Comparatively, the number of mixed multifurcating 
trees grows at a hyper-exponentially faster rate, as can be seen in \cref{fig:plot_trees}. This 
motivates a need for an alternative method to exhaustive enumeration of all $n$-species trees.

\section{Maximum Parsimony for trees with labelled leaves}\label{sec:small_pars}

According to the parsimony criterion, we seek a tree that explains divergence
of species with the fewest number of evolutionary events. As such, we seek to identify
a tree with $n$ labelled nodes and fitted unlabelled nodes such that the tree cost, which is the sum
of edge costs and therefore the total number of mutations, is minimized. This problem can be
broken into two subproblems.
\extabs{{\bf Small parsimony problem (SPP):} Given a tree $\tau$ with $n$ species nodes and 
a specified topology, compute its MP-cost. 
{\bf Large parsimony problem (LPP):} Given a set of $n$ species, find the tree(s) with the 
minimum MP-cost among all possible tree topologies. Such tree(s) is/are called the {\it most parsimonious tree(s)} ({\it MP-trees}).
}

\subsection{Hartigan's algorithm}\label{subsec:hartigan}

Hartigan's algorithm provides a\fullversion{more} powerful framework for calculating best fits of a 
given tree. It solves the SPP for multifurcating rooted trees with $n$ labelled leaves 
\cite{Har73}. The bottom-up procedure of Hartigan's algorithm processes every unlabelled 
internal node $u_i, 1 \le i \le n-2$ that has children $v_i, i \geq 2.$ The procedure 
recursively calculates upper $VU(u)_i$ and lower $VL(u)_i$ sets for every character of every
unlabelled node $u$ as 
\extabs{follows (theorem 2 in \cite{Har73}):

If \swallow{$n(u)$ is the number of progeny of node $u$, }$k(A)$ the number of times a value $A$ occurs in the sets $VU(u)_i$ of every child $v$ of $u$, and $K = \max{k(A)}$, then 
\begin{enumerate}
\item $VU(u)_i = \{ A | k(A) = K\}$
\item $VL(u)_i = \{ A | k(A) = K-1\}$
\end{enumerate}
}
\fullversion{shown in \cref{algo:hartigan_up}.

\begin{algorithm}[t]
\caption{Hartigan's bottom up procedure}
\label{algo:hartigan_up}
\begin{algorithmic}[1]
\Procedure{Hartigan}{}
\State {count $= 0$;}
\For {each $C_i, 1 \le i \le m$}
\If{$v_i$ is a leaf}
\State{Initialize $VU(v)_i$;}
\State{$VL(v)_i=\emptyset$;}
\State {count $+= 0$;}
\Else
\State {$num(v_i)=0$;}
\For {each state $x_j \in C_i$} 
\State {$num(x_j)=0$;}
\EndFor
\For {each child $v_i$}
\If{$x_j \in VU(v)_i$}
\State {$num(x_j)_i ++$;}
\EndIf
\State {$num(v_i)_i ++$;}
\EndFor
\State {$K = max(num(x_j))$;}
\State {$VU(v)_i = \{x_j|num(x_j) == K\}$;}
\State {$VL(v)_i = \{x_j|num(x_j) == K-1\}$;}
\State {count += $Hartigan(v_i) + num(v_i) - K$;}
\EndIf
\EndFor
\Return {count;}
\EndProcedure
\end{algorithmic}
\end{algorithm}
}

Hartigan's top down refinement allows the computation of optimal assignments to each node. For any character $i$ of the root node, selecting any of the candidate states from its root set would yield a most parsimonious labelling. The algorithm then proceeds to compute the root sets of characters of internal nodes $v$ using
\extabs{the following result (theorem 3 in \cite{Har73}):

For $v$ child of $u$:
\begin{enumerate}
\item If $VV(u)_i \subseteq VU(v)_i$, then $VV(v)_i = VV(u)_i$
\item Otherwise, $VV(v)_i = VU(v)_i \cup (VV(u)_i \cap VL(v)_i)$
\end{enumerate}
}
\fullversion{\cref{algo:hartigan_down}.

\begin{algorithm}[b]
\caption{Hartigan's top down fitting}
\label{algo:hartigan_down}
\begin{algorithmic}[1]
\For {each $C_i, 1 \le i \le m$}
\If {$VV(u)_i \subseteq VU(v)_i$\;}
\State {$VV(v)_i=VV(u)_i$\;}
\Else 
\State {$VV(v)_i = VU(v)_i \cup (VV(u)_i \cap VL(v)_i)$\;}
\EndIf
\EndFor
\end{algorithmic}
\end{algorithm}
}

\begin{figure}[bt]
\begin{center}
\includegraphics[width=9cm]{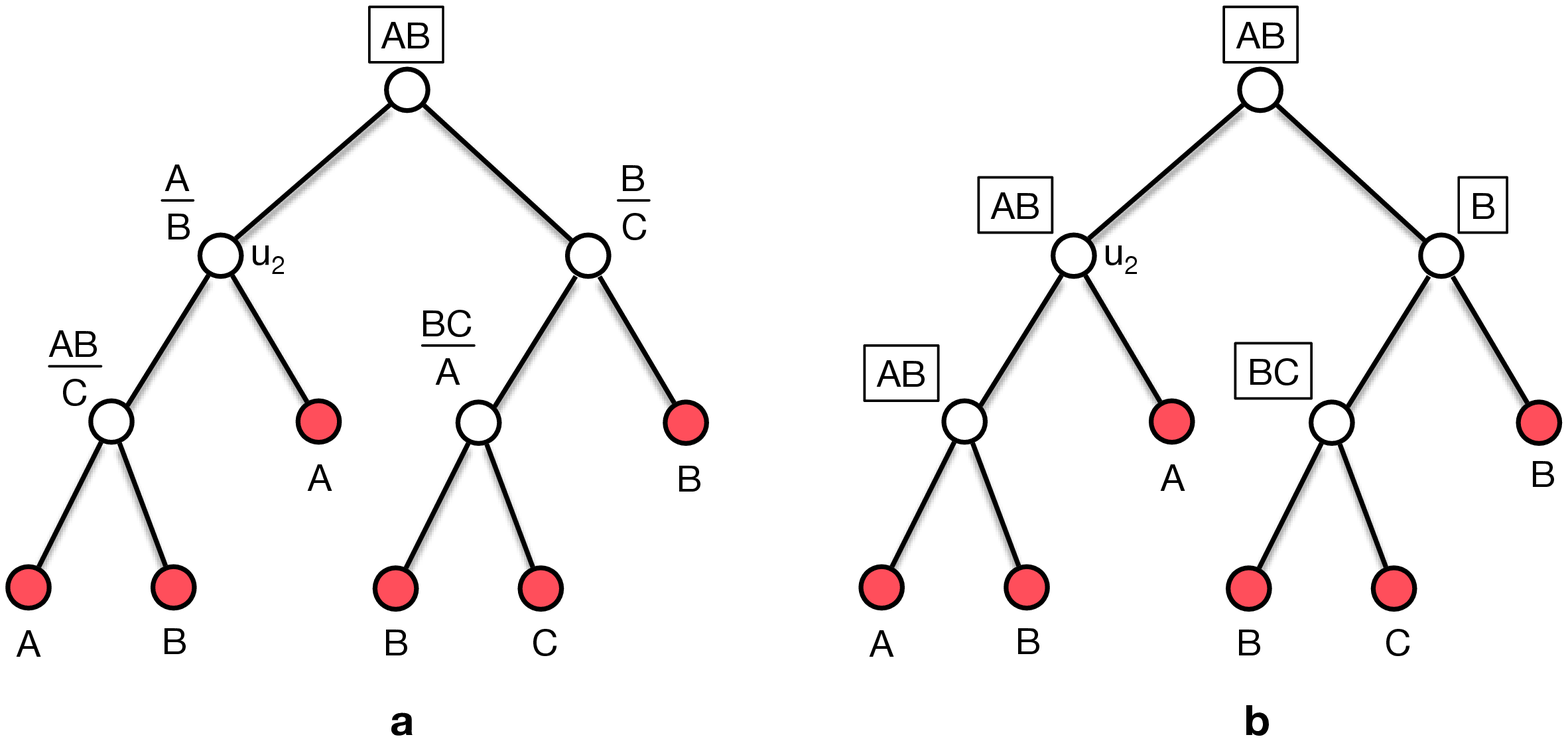}
\end{center}
\caption{Hartigan's algorithm: (a) Computation
of upper $VU$ and lower $VL$ sets of all unlabelled nodes, as well as root $VV$ set of root node, after bottom-up step. (b) Computation of root sets $VV$ of all unlabelled nodes after top-down step.}
\label{fig:hartigan}
\end{figure}

By storing all optimal and next-to-optimal values in sets $VU(u)$ and $VL(u)$ respectively, 
and by computing $VV(u)$, Hartigan's algorithm can be used to find all co-optimal solutions 
to the SPP. An example of Hartigan's algorithm can be seen in \cref{fig:hartigan}. 
\fullversion{In this example we can observe the same tree as in \cref{fig:fitch}, 
where state $B$ is now included in the root set of node $u_2$.
}

\fullversion{
It is important to note that Fitch's and Hartigan's algorithms are restricted to unit cost character mutations. Other algorithms exist for weighted costs, such as Sankoff's algorithm \cite{San75}, which allows for edge independent, arbitrary metric costs that may vary for each evolutionary event.
}

\subsection{Unrooting trees}\label{subsec:unroot}

Tree enumeration for the LPP on rooted full binary trees involves the systematic generation
of cubic trees, for which MP-costs are computed by arbitrarily rooting the
trees. To maintain bifurcation, a root can be added to a tree by replacing an edge $(v_1, v_2)$ with a new unlabelled
root node $r$ and two edges $(r, v_1)$ and $(r, v_2)$. It is evident that the cost of
the new tree will remain unaltered, since the root node can be assigned the same root set and value
as one of either $v_1$ or $v_2$.

Conversely, the following theorem also holds true:
\begin{theorem}
\label{theorem:unrooting}
Removing the root of a binary tree, as well as any unlabelled node of degree 2, does not
change the MP-cost of the tree.
\end{theorem}
\begin{proof}
Hartigan's algorithm on a rooted binary tree
computes the root sets of all internal nodes, including the root set $VV(r)$ of the root node $r$. Let $VV(x)$ and
$VV(y)$ be the corresponding root sets of the root's children $x$ and $y$. Any assignment
of a state $S_i \in VV(r)_j$ to the character $C_j$ of the root node will result in
\begin{enumerate}
\item A cost of $0$ for mutating this character from the root to both children,
if $VV(r)_j \in (VV(x)_j \cap VV(y)_j)$ or
\item A cost of $1$ otherwise (when $VV(r)_j \in (VV(x)_j \cup VV(y)_j$)).
\end{enumerate}
Removing the root and connecting nodes $x$ and $y$ directly with an edge will not cause
an increase to the MP-cost of the tree, as the same assignments that would minimize
the edge costs between the root and its children will now be maintained on the
edge $(x, y)$, meaning $0$ for each character $j$ whose state does not 
mutate ($VV(x)_j \cap VV(y)_j \ne \emptyset$), and $1$ when the state mutates.
\end{proof}
Therefore, cubic trees with $n$ labelled leaves share the same MP-cost
with binary rooted counterparts (not necessarily full).

\section{Towards a compact most parsimonious tree}\label{sec:compact}

Our ultimate goal is to find the most compact parsimonious $n$-species trees. 
To solve this problem, in this section we will demonstrate that it is sufficient 
to find the cubic $n$-species MP-trees and contract them. 
Towards that goal we will prove that most compact $n$-species 
MP-tree cannot have a lower cost fit than the cubic $n$-species 
MP-tree. To prove this claim we will utilize the following lemmas:

\begin{lemma}
\label{lemma:labels_on_leaves}
An $n$-species MP-tree with labelled internal nodes cannot have a 
lower cost than an $n$-species MP-tree with $n$ labelled leaves.
\end{lemma}
\begin{proof}
We will prove by construction, while maintaining the invariant of lowest tree cost. 
Consider an $n$-species MP-tree with labelled
internal nodes. Let $u_i$ be one of these nodes. Let $(u_i, v)$ be an edge
connecting $u_i$ with another node $v$. We will create a new internal node $u_k$
with the same root set as $u_i$, meaning $VV(u_k) = VV(u_i)$. We will then
remove the $(u_i, v)$ edge and connect $u_k$ to $u_i$ and $v$ with two
edges $(u_i, u_k)$ and $(u_k, v)$. We will then create a new leaf node $u_l$ with
$VV(u_l) = VV(u_i)$ and connect it to node $u_k$ with an edge $(u_k, u_l)$. Finally we
will move the label from $u_i$ to $u_l$, effectively removing a labelled
internal node and creating a labelled leaf. The construction is shown in \cref{fig:labels_on_leaves}.

\begin{figure}[bt]
\begin{center}
\includegraphics[width=9cm]{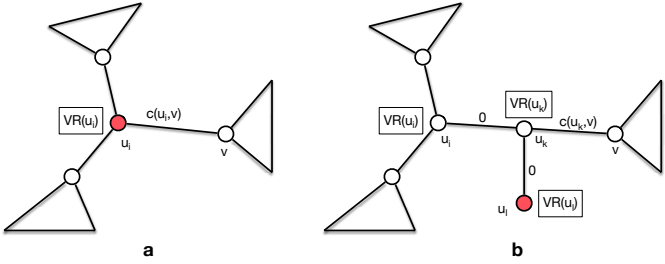}
\end{center}
\caption{Moving an internal labelled node to a leaf while maintaining tree cost}
\label{fig:labels_on_leaves}
\end{figure}

The tree cost remains unchanged during these operations, since edge $(u_k, v)$
will have the same potential cost (for the same fit of $v$) as edge $(u_i, v)$ had,
where the other new edges $(u_i, u_k)$ and $(u_k, u_l)$ will have potential costs of $0$,
since they connect nodes with the same single-fit root sets. We can repeat this
process independently on every internal labelled node, until the only labelled
nodes are leaves, while the MP-cost of the tree remains the same.
\end{proof}

\begin{lemma}
\label{lemma:split}
In leaf-labelled trees, a multifurcating $n$-species MP-tree cannot have a lower cost fit than an $n$-species cubic MP-tree.
\end{lemma}
\begin{proof}
We will prove this lemma by construction, once again without modifying the MP-tree cost. 
A multifurcating tree has two types of nodes that do not appear in a cubic tree, nodes 
of degree $2$ and nodes of degrees $\ge 4$. We have seen how to remove unlabelled 
nodes of degree 2 in \cref{theorem:unrooting} without increasing the MP-tree cost. 
To remove tree nodes with degrees greater than $3$ we will introduce a split operation 
that will create a new node, reduce the degree of an existing node by $1$, and 
conserve the tree cost.

\begin{figure}[bt]
\begin{center}
\includegraphics[width=9cm]{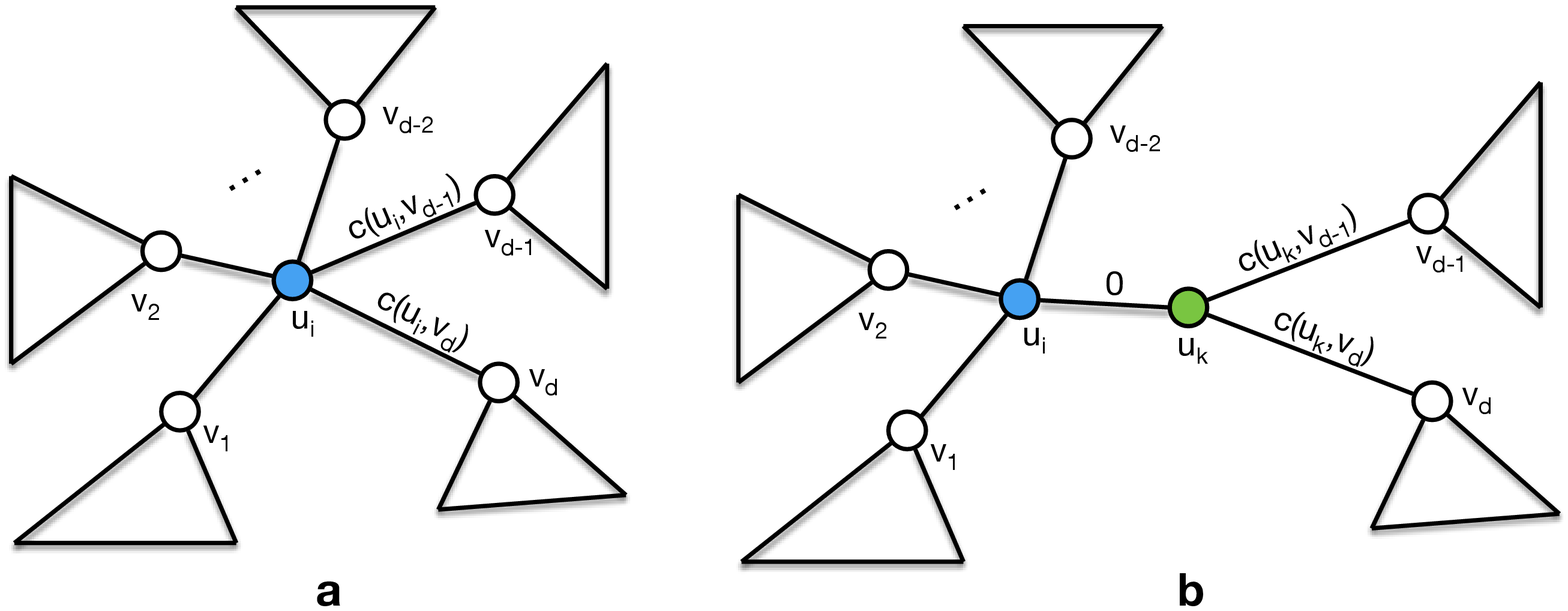}
\end{center}
\caption{Node split to reduce degree of a node while maintaining tree cost}
\label{fig:split}
\end{figure}

Consider a node $u_i$ with degree $d > 3$. Node $u_i$ will be adjacent to $d$ other 
nodes $v_1, v_2, \dots, v_d$. We will create a new unlabelled node $u_k$ with the same 
root set as $u_i$, which we will connect to $u_i$. Then we will disconnect nodes 
$v_{d-1}$ and $v_d$ from $u_i$, and connect them to $u_k$. The modified node $u_i$ is now connected to 
nodes $v_1, v_2, \dots, v_{d-2}, u_k$ and has degree $d-1$, where node $u_k$ is 
adjacent to $u_i, v_{d-1}$ and $v_d$, and has degree $3$. The degrees of all other nodes 
are unchanged. The MP-tree cost remains the same, as new edge $(u_i, u_k)$ has a potential cost of 
0 with an original tree best fit, and removed edges $(u_i, v_{d-1})$ and $(u_i, v_d)$ carry 
the same potential cost with added 
edges $(u_k, v_{d-1})$ and $(u_k, v_d)$ respectively. The split operation is shown in 
\cref{fig:split}.

Repeating the split operation on all nodes with degrees $\ge 4$ until their degrees are reduced to $3$ 
will produce a cubic
tree with the same MP-cost as the original multifurcating tree.
\end{proof}

\begin{lemma}
\label{lemma:no_unlabel}
Unlabelled nodes with degrees $d < 3$ can be removed from an $n$-species tree without increasing its MP-cost.
\end{lemma}
\begin{proof}
We have seen how to remove unlabelled nodes of degree $2$ in \cref{theorem:unrooting} 
without increasing the tree MP-cost.
To remove an unlabelled leaf $v$, we can notice that its incident edge can always have a cost of $0$ for any given fit, since we can always set $VV(v) = VV(u)$, where $u$ is the single neighbor of $v$. As such, removal of $v$ and its incident edge does not increase the tree cost.
\end{proof}

\begin{lemma}
\label{lemma:unlabel_bound}
A most compact parsimonious $n$-species tree will have at most $n-2$ unlabelled nodes.
\end{lemma}
\begin{proof}
Based on \cref{lemma:no_unlabel}, all leaves of a most compact MP-tree
will be labelled. Thus, such a tree will have $n$ leaves.
Assume to the contrary of our stated lemma that a most compact $n$-species MP-tree 
has $k \ge n-1$ internal nodes, all of which have degrees $\ge 3$, 
as per \cref{lemma:no_unlabel}. Then the total number of nodes 
of the tree is $n + k$. A tree with $n+k$ nodes has $n+k-1$ edges. 
The sum of the node degrees then will be $2n+2k-2$, since every edge contributes $2$ to the total sum.

The sum of the degrees of the $n$ leaves is $n$, which means that the sum of degrees of the internal nodes $S = n+2k-2$. Since every internal node has a degree $\ge 3$, the $k$ internal nodes will have a sum of degrees 
$S \ge 3k \Leftrightarrow n+2k-2 \ge 3k \Leftrightarrow k \le n-2$, which contradicts our assumption $k \ge n-1$.
\end{proof}

Now we can proceed with the proof of our main theorem:

\begin{theorem}
\label{theorem:main}
The most compact $n$-species MP-tree cannot have a lower cost fit than the $n$-species cubic MP-tree.
\end{theorem}
\begin{proof}
Assume to the contrary that there exists a tree $\tau_c$ on $n$ species $S$ that has a lower cost than 
the cubic MP-tree $\tau$ on $S$. Using the construction in \cref{lemma:labels_on_leaves}
we can move all labelled internal nodes to leaves without increasing the MP-cost of $\tau_c$. 
Based on \cref{lemma:unlabel_bound} we could remove all unlabelled nodes with 
degrees $\le 2$ without altering the MP-cost of $\tau_c$ as well. 
Now $\tau_c$ has only nodes with degree $1$ or degree $\ge 3$. Using the construct of \cref{lemma:split} we
can convert $\tau_c$ to a cubic tree, by successively splitting nodes of degree higher than 3, again without
affecting the MP-cost of $\tau_c$. The resulting tree is cubic, has all species in $S$ associated with leaves,
and a lower cost than $\tau$\fullversion{, which contradicts the optimality of $\tau$}.
\end{proof}

\Cref{theorem:main} enables us to build the most compact MP-tree
without enumerating all $n$-species trees, but only cubic trees with $n$ species. It also supplies us
with a systematic procedure to create the most compact parsimonious tree by reversing the process described in
\cref{theorem:main}. Starting with the $n$-species cubic MP-trees, we can contract edges
with $0$ min-cost, effectively reversing the split operation. But which is the right order
to contract edges, so that we can produce the most compact parsimonious tree? The relation 
$R : V\to V : (u, v) \in R \iff md(u, v) = 0$ is not transitive, and edge contraction order can matter.
Therefore we will consider all possible orders of edge contractions.

\begin{lemma}
\label{lemma:root_set}
The root tuple $VV(v)$ of a node $v$ is independent of the placement of the root of the tree and 
its character sets are maximal.
\end{lemma}
\begin{proof}
$VV(v)$ indicates the tuple of maximal sets of states that can be assigned to corresponding characters
of $v$ in a most parsimonious fit. These are computed by the top-down procedure of Hartigan's algorithm
\fullversion{\cref{algo:hartigan_down}}, the correctness of which is proven in theorem 3 of \cite{Har73}.
\end{proof}

\begin{corollary}
\label{root_set_upper_set}
$VV(v) = VU(v)$ when $v$ is placed on the root.
\end{corollary}

The following lemma will help us prove the correctness of our contraction algorithm.

\begin{lemma}
\label{lemma:contraction}
Only edges with $0$ min-cost can be contracted without increasing tree cost.
\end{lemma}
\begin{proof}
Assume to the contrary that we can contract an edge with min-cost $>0$ where the contracted tree
$\tau_c$ has the same MP-cost $MM(\tau)$ as the initial MP-tree $\tau$. Let $(u,v)$ be such an edge. 
Then $\exists i, 1 \le i \le m: VV(v)_i \cap VV(u)_i = \emptyset$.

Let $w$ be the new node created once edge $(u,v)$ is contracted. A value $V(w)$ with
$V(w)_i = x \in VV(u)_i$ (we select $u$ without loss of generality) 
would set $MM(\tau_c) > MM(\tau)$. To see that, let us root $\tau$
at $u$. Clearly $VV(u)_i \not\subseteq VV(v)_i$, which means, based on \fullversion{\cref{algo:hartigan_down}}\extabs{theorem 2 of \cite{Har73}} and \cref{lemma:root_set}, that 
$VV(v)_i = VU(v)_i \cup (VV(u)_i \cap VL(v)_i)$. But then 
\fullversion{\begin{itemize}
\item }$x \not\in VU(v)_i$, since $x \in VU(v)_i \implies x \in VV(v)_i \implies VV(v)_i \cap VV(u)_i \ne \emptyset$.
\fullversion{\item }Also $x \not\in VL(v)_i$, since $x \in VL(v)_i \implies x \in VV(u)_i \cap VL(v)_i \implies x \in VV(v)_i
\implies VV(v)_i \cap VV(u)_i \ne \emptyset$.
\fullversion{\end{itemize}}
Thus an assignment of $x$ to $V(v)_i$ in $\tau$ would increase the cost of the subtree
rooted at $v$ by $2$ more than any other assignment from $VV(v)_i = VU(v)_i$. Even with the gain
of one mutation from the contraction of edge $(u,v)$, $MM(\tau_c) > MM(\tau)$, which is a contradiction.
\end{proof}

\begin{corollary}
\label{corollary:contracted_root_set}
After an edge $(u,v)$ contraction, the new node $w$ will have root set 
$VV(w): \forall_i 1\le i\le m, VV(w)_i = VV(u)_i \cap VV(v)_i$.
\end{corollary}
\begin{proof}
The proof follows from the same process as in \cref{lemma:contraction}.
\end{proof}

Our algorithm for contracting a cubic tree to the most compact tree on $n$ species with the same cost
is described in \cref{algo:compact}.

\begin{algorithm}
\caption{Tree contraction algorithm}
\label{algo:compact}
\begin{algorithmic}[1]
\Loop { For each $0$-min-cost edge}
\State {Contract edge $(u,v)$, remove $u, v$ and create new node $w$}
\State {Update root set of w: $\forall_i 1\le i\le m VV(w)_i = VV(u)_i \cap VV(v)_i$}
\State {Root tree on w and run DFS to:}
\State {(a) Update root sets of all unlabelled tree nodes and }
\State {(b) Update list of $0$-min-cost edges}
\EndLoop
\end{algorithmic}
\end{algorithm}

\begin{theorem}
\label{theorem:compact}
\Cref{algo:compact} yields the most compact parsimonious tree
\end{theorem}
\begin{proof}

\fullversion{
The contraction of an edge $(u, v)$ creates
a new vertex $w$, whose root set is computed as follows: 
$VV(w)_i = VV(u)_i \cap VV(v)_i, 1 \le i \le m$.

The root sets of all remaining unlabelled nodes in the tree
are updated recursively via a DFS traversal starting at node $w$.
For every node $u$ whose initial root set was $VV(pu)$, 
whose parent's previous root set was $VV(pp)$
and whose parent's current root set is $VV(cp)$, the new root set $VV(u)$ is
calculated as follows: 
$VV(u)_i = (VV(pu) \setminus VV(pp)_i) \cup (VV(pu)_i \cap VV(cp)), 1 \le i \le m$.
}

The proof of correctness of \cref{algo:compact} follows from the reversal of the conversion of the most compact MP-tree to a cubic MP-tree. We are exhaustively enumerating all $n$-species cubic trees and, for the most parsimonious of them, we are considering all possible orders of edge contractions. Contracting edges reverses the node split operation that was utilized in \cref{theorem:main}.
\end{proof}

The initial cubic tree (before any contractions) has $n$ labelled and $n-2$ unlabelled nodes, therefore
$2n-2$ nodes and $2n-3$ edges. The minimum number
of nodes a compact tree can have is $n$, so the maximum number of consecutive contractions that can be
performed is $n-2$. In the worst case our algorithm can iterate $(2n-3)(2n-4)\cdot \dots \cdot n = \binom{2n-3}{n-2}$ times.
Each iteration involves a DFS traversal that takes linear time as a function of the size of the tree.
Therefore the worst-case time complexity of the edge contraction algorithm is hyperexponential. 
Previous work in {\it tree refinement}, where maximum parsimony is pursued by contracting edges in trees, has
shown that the tree refinement problem is NP-hard \cite{Bon98}, indicating that our problem may not have 
efficient solutions in the worst case without bounding the values of any parameters.
On average we would expect the edge contraction
algorithm to be much more efficient, as the probability of contracting a tree edge
decreases exponentially as a function of the number of characters examined, assuming character independence.

\swallow{
Would it be possible to derive a more compact tree by merging non-adjacent nodes? The answer to this question
is negative, as the following lemma demonstrates.

\begin{theorem}
\label{theorem:neighbor_merge}
We only need to consider edge contractions for promotions.
\end{theorem}
\begin{proof}
Let us consider the possible promotion of a node $u$ to an internal node $v$. Node $u$
can either be labelled or unlabelled, where $v$ has to be unlabelled, otherwise
the promotion would not be feasible. After the promotion, the node 
\end{proof}

The order of merge operations does not matter? Order may matter. If an internal node u 
can be merged with two neighbors $u_1$ and $u_2$, as well as a leaf v, and u, v, and 
$u_1$ have a non-empty intersection, but u, v, and $u_2$ do not, then merging u with 
$u_2$ precludes the leaf and resulting node merge, where all three u, $u_1$, and v 
could be merged in one node.
Maybe search for longest paths of collectively non-empty intersecting nodes? We could 
possibly find that by comparing all pairs of nodes in a subtree
}

\section{Experimental Results}\label{sec:results}

\begin{table*}[!t]
\renewcommand{\arraystretch}{1.3}
\caption{Mixed Tree Enumeration vs Cubic MP-Tree Contraction}
\label{table:results}
\centering
\begin{tabular}{c||c|c|c|c|c|c}
\hline
\bfseries Number of & \bfseries MTEA     & \bfseries CTEECA     & \bfseries Compact Mixed & \bfseries Cubic         & \bfseries Contracted Cubic & \bfseries Number of \\
\bfseries Species   & \bfseries Time (ms)& \bfseries Time (ms) & \bfseries MP-Trees (\#) & \bfseries MP-Trees (\#) & \bfseries MP-Trees (\#)     & \bfseries Contractions\\
\hline\hline
4 & 10       & 8       &  1.1 & 1.1  &  3      &  1.4\\
5 & 36       & 10      &  1.6 & 1.9  &  5.4    &  1.7\\
6 & 191      & 43      &  2   & 2.7  &  13.4   &  1.8\\
7 & 796      & 167     &  3.5 & 4.4  &  162    &  2.3\\
8 & 8540     & 811     &  2.5 & 4    &  30.2   &  2.4\\
9 & 128242   & 10458   &  4.8 & 10.5 &  328.8  &  2.8\\
10& 865839   & 139922  &  5.2 & 10.1 &  752    &  3.1\\
11& 3831436  & 1778987 &  3.9 & 18.7 &  1716.8 &  3.5\\
\hline
\end{tabular}
\end{table*}

We implemented two branch-and-bound algorithms to identify the most compact MP-trees for $n$ species.
The first algorithm, the Mixed Tree Enumeration Algorithm (MTEA), exhaustively enumerates all mixed trees
to identify the most compact MP-trees. The second algorithm, the Cubic Tree Enumeration
and Edge Contraction Algorithm (CTEECA), exhaustively
enumerates all cubic trees to find the MP-trees, on which it applies our edge contraction method to
identify the most compact MP-trees. 

We run these two algorithms
on a dataset of viral sequences from the genes {\it lef-8} and {\it ac22} of the 
{\it Baculoviridae} family, analyzed in \cite{Her04}. A multiple alignment of these
sequences was downloaded from TreeBASE \cite{Pie02}, and the first 30 characters of
each taxon were used. We excluded 4 taxa for which the first 30 characters were not
known.

We run our two algorithms to find most compact MP-trees for $n$ species with $4 \le n \le 11$. 
The $n$ species were selected at random from the 35 available sequences, and for each value of $n$
we run 10 seperate randomized experiments, averaging the results.
All experiments were run on a desktop computer with an Intel i7-4820k processor running at
3.7Ghz with 16 GB of RAM, an amount adequate for
all data to be stored in memory once the input sequences were imported from the solid state drive.

The results of our experiments are shown in \cref{table:results}.  Execution times displayed 
include the time needed to enumerate and score mixed trees, or the time needed to enumerate, 
score, and contract cubic trees. Running times varied significantly among trials for any given 
$n$, due to the nature of branch and bound algorithms; in some trials, low scoring trees 
were found earlier on in the enumeration, allowing for more efficient pruning of the search
space. The number of compact mixed trees reported is the mean number of most parsimonious 
mixed trees that had the fewest number of nodes. The number of most compact trees from the 
cubic enumeration includes all possible most compact trees generated by contracting edges 
in the most parsimonious cubic trees. This comparatively high number includes possible 
duplicate trees as well as possible rerootings of the same tree. The number of contractions 
is the mean number of zero min-cost edges that were contracted in the most parsimonious 
cubic trees. This shows the average difference in the number of nodes between the most compact 
trees and the cubic trees from which they were generated.
Our results experimentally demonstrate that the CTEECA outperforms the MXEA by at least an order of magnitude on
reasonable biological datasets, with similar behavior observed on datasets from the
domain of phylogenetic stemmatics.

The programs used to perform these experiments were written in the Java programming language and 
the documented source code be downloaded at:
\url{https://github.com/ottj3/phylotreecontract}.

\section{Conclusion}\label{sec:conclusion}

In this work we have established a novel connection between mixed MP-trees
and cubic MP-trees, and shown a mapping from the cubic MP-trees to the most 
compact mixed MP-trees, enabling more efficient algorithms for live phylogeny
with polytomies.
We have designed and implemented an efficient optimal algorithm to generate the most compact MP-trees
for $n$ species by enumerating all cubic $n$-species trees, finding the most
parsimonious ones, and optimally contracting them. Although contraction requires
potentially hyper-exponential time as a function of the number of species, the running time
of our algorithm is superior to the enumeration of all multifurcating trees
with $n$ species, even in the worst case. On average we expected the contraction
algorithm to be comparatively very efficient, an expectation that was confirmed experimentally.
Furthermore,
cubic tree enumeration has been refined in several existing phylogenetic
software suites for many years 
\cite{Fel89,Swo03}, and a large number of heuristics, approximations, and parallel
algorithms have been developed and used effectively to speed up 
enumeration \cite{Bon98,Gol02,Yan03,Bad06,Sri08,Goe08,Alo10,Whi11}, 
advancements of which our edge contraction algorithm can easily take advantage to further improve its efficiency.

It is our hope that our theoreticaly advances in the understanding of maximum live parsimony
with polytomies and our optimal algorithms for identifying the most compact MP-trees
for $n$ species -- providing the ability to handle polytomies and input species on internal nodes natively --
will enhance studies and enable new advances in evolutionary virology, 
paleontology, linguistics, and phylogenetic stemmatics.

\section*{Acknowledgments}
This work has been supported by NSF Grant CCF-1418874 and The College of New Jersey
Mentored Undergraduate Summer Experience (MUSE) program.

\ifCLASSOPTIONcaptionsoff
  \newpage
\fi

\bibliographystyle{IEEEtran}

\bibliography{paper.bib}

\begin{thebibliography}{10}
\providecommand{\url}[1]{#1}
\csname url@samestyle\endcsname
\providecommand{\newblock}{\relax}
\providecommand{\bibinfo}[2]{#2}
\providecommand{\BIBentrySTDinterwordspacing}{\spaceskip=0pt\relax}
\providecommand{\BIBentryALTinterwordstretchfactor}{4}
\providecommand{\BIBentryALTinterwordspacing}{\spaceskip=\fontdimen2\font plus
\BIBentryALTinterwordstretchfactor\fontdimen3\font minus
  \fontdimen4\font\relax}
\providecommand{\BIBforeignlanguage}[2]{{%
\expandafter\ifx\csname l@#1\endcsname\relax
\typeout{** WARNING: IEEEtran.bst: No hyphenation pattern has been}%
\typeout{** loaded for the language `#1'. Using the pattern for}%
\typeout{** the default language instead.}%
\else
\language=\csname l@#1\endcsname
\fi
#2}}
\providecommand{\BIBdecl}{\relax}
\BIBdecl

\bibitem{Ron01}
\BIBentryALTinterwordspacing
J.~Huelsenbeck and F.~Ronquist, ``Mrbayes: Bayesian inference of phylogenetic
  trees,'' \emph{Bioinformatics}, vol.~17, no.~8, pp. 754--755, 2001. [Online].
  Available: \url{http://www.ncbi.nlm.nih.gov/pubmed/11524383}
\BIBentrySTDinterwordspacing

\bibitem{Fel81}
J.~Felsenstein, ``{Evolutionary trees from DNA sequences: A maximum likelihood
  approach},'' \emph{Journal of Molecular Evolution}, vol.~17, no.~6, pp.
  368--376, 1981.

\bibitem{Hil92}
D.~Hillis, J.~Bull, M.~White, M.~Badgett, and I.~Molineux, ``Experimental
  phylogenetics: generation of a known phylogeny,'' \emph{Science}, vol. 255,
  no. 5044, pp. 589--592, 1992.

\bibitem{Zhu98}
T.~Zhu, B.~Korber, A.~Nahmias, E.~Hooper, P.~Sharp, and D.~Ho, ``An african
  hiv-1 sequence from 1959 and implications for the origin of the epidemic,''
  \emph{Nature}, vol. 391, no. 6667, pp. 594--597, 1998.

\bibitem{Bus99}
\BIBentryALTinterwordspacing
R.~Bush, C.~Bender, K.~Subbarao, N.~Cox, and W.~Fitch, ``{Predicting the
  evolution of human influenza A},'' \emph{{Science}}, vol. {286}, no. {5446},
  pp. {1921--1925}, {1999}. [Online]. Available:
  \url{http://www.ncbi.nlm.nih.gov/pubmed/10583948}
\BIBentrySTDinterwordspacing

\bibitem{Wil03}
\BIBentryALTinterwordspacing
G.~D. Wilson and G.~D. Edgecombe, ``The triassic isopod protamphisopus
  wianamattensis (chilton) and comparison with extant taxa (crustacea,
  phreatoicidea),'' \emph{Journal of Paleontology}, vol.~77, no.~3, pp.
  454--470, 2003. [Online]. Available:
  \url{http://www.jstor.org/stable/pdf/4094794.pdf?seq=1#page_scan_tab_contents}
\BIBentrySTDinterwordspacing

\bibitem{Sal02}
N.~Salamin, T.~R. Hodkinson, and V.~Savolainen, ``{Building supertrees: an
  empirical assessment using the grass family (Poaceae).}'' \emph{Systematic
  biology}, vol.~51, no.~1, pp. 136--150, 2002.

\bibitem{Swo09}
\BIBentryALTinterwordspacing
D.~L. Swofford and J.~Sullivan, ``{Phylogeny inference based on parsimony and
  other methods using PAUP},'' in \emph{The Phylogenetic Handbook}, 2nd~ed.,
  P.~Lemey, M.~Salemi, and A.-M. Vandamme, Eds.\hskip 1em plus 0.5em minus
  0.4em\relax Cambridge University Press, 2009, pp. 267--312, cambridge Books
  Online. [Online]. Available:
  \url{http://dx.doi.org/10.1017/CBO9780511819049.010}
\BIBentrySTDinterwordspacing

\bibitem{Tel13}
\BIBentryALTinterwordspacing
G.~P. Telles, N.~F. Almeida, R.~Minghim, and M.~E. M.~T. Walter, ``{Live
  phylogeny.}'' \emph{Journal of computational biology : a journal of
  computational molecular cell biology}, vol.~20, no.~1, pp. 30--7, 2013.
  [Online]. Available: \url{http://www.ncbi.nlm.nih.gov/pubmed/23294270}
\BIBentrySTDinterwordspacing

\bibitem{Car13}
\BIBentryALTinterwordspacing
S.~A. Carroll, J.~S. Towner, T.~K. Sealy, L.~K. McMullan, M.~L. Khristova,
  F.~J. Burt, R.~Swanepoel, P.~E. Rollin, and S.~T. Nichol, ``Molecular
  evolution of viruses of the family filoviridae based on 97 whole-genome
  sequences,'' \emph{Journal of Virology}, vol.~87, no.~5, pp. 2608--2616,
  2013. [Online]. Available:
  \url{http://jvi.asm.org/content/87/5/2608.abstract}
\BIBentrySTDinterwordspacing

\bibitem{Bue09}
P.~Buendia and G.~Narasimhan, ``{Serial evolutionary networks of within-patient
  HIV-1 sequences reveal patterns of evolution of X4 strains.}'' \emph{BMC
  Systems Biology}, vol.~3, p.~62, 2009.

\bibitem{Foo96}
\BIBentryALTinterwordspacing
M.~Foote, ``On the probability of ancestors in the fossil record,''
  \emph{Paleobiology}, vol.~22, pp. 141--151, 1996. [Online]. Available:
  \url{http://journals.cambridge.org/article_S0094837300016146}
\BIBentrySTDinterwordspacing

\bibitem{Mie01}
\BIBentryALTinterwordspacing
P.~Mierzejewski, ``A new graptolite, intermediate between the tuboidea and the.
  camaroidea,'' \emph{Acta Palaeontologica Polonica}, vol.~46, no.~3, pp.
  367--376, 2001. [Online]. Available:
  \url{https://www.app.pan.pl/archive/published/app46/app46-367.pdf}
\BIBentrySTDinterwordspacing

\bibitem{Urb98}
A.~Urbanek, ``\BIBforeignlanguage{English}{Oligophyly and evolutionary
  parallelism: A case study of silurian graptolites},''
  \emph{\BIBforeignlanguage{English}{Acta Palaeontologica Polonica}}, vol.~43,
  no.~4, pp. 549--572, 1998.

\bibitem{Bap13}
\BIBentryALTinterwordspacing
D.~W. Bapst, ``{paleotree: an R package for paleontological and phylogenetic
  analyses of evolution},'' \emph{Methods in Ecology and Evolution}, vol.~3,
  no.~5, pp. 803--807, 2012. [Online]. Available:
  \url{http://dx.doi.org/10.1111/j.2041-210X.2012.00223.x}
\BIBentrySTDinterwordspacing

\bibitem{Wal13}
\BIBentryALTinterwordspacing
A.~F.~Y. Poon, L.~W. Walker, H.~Murray, R.~M. McCloskey, P.~R. Harrigan, and
  R.~H. Liang, ``{Mapping the Shapes of Phylogenetic Trees from Human and
  Zoonotic RNA Viruses},'' \emph{PLoS ONE}, vol.~8, no.~11, pp. 78--122, 2013.
  [Online]. Available: \url{http://dx.doi.org/10.1371%2Fjournal.pone.0078122}
\BIBentrySTDinterwordspacing

\bibitem{Gol02}
P.~A. Goloboff, ``{Optimization of polytomies: state set and parallel
  operations.}'' \emph{Molecular phylogenetics and evolution}, vol.~22, no.~2,
  pp. 269--275, 2002.

\bibitem{Gra82}
R.~L. Graham and L.~R. Foulds, ``{Unlikelihood that minimal phylogenies for a
  realistic biological study can be constructed in reasonable computational
  time},'' \emph{Mathematical Biosciences}, vol.~60, no.~2, pp. 133--142, 1982.

\bibitem{Day83}
W.~H.~E. Day, ``{Computationally difficult parsimony problems in phylogenetic
  systematics},'' \emph{Journal of Theoretical Biology}, vol. 103, no.~3, pp.
  429--438, 1983.

\bibitem{Day86}
W.~H.~E. Day, D.~S. Johnson, and D.~Sankoff, ``{The computational complexity of
  inferring rooted phylogenies by parsimony},'' \emph{Mathematical
  Biosciences}, vol.~81, no.~1, pp. 33--42, 1986.

\bibitem{Car14}
A.~Carmel, N.~Musa-Lempel, D.~Tsur, and M.~Ziv-Ukelson, ``{The worst case
  complexity of maximum parsimony},'' in \emph{Lecture Notes in Computer
  Science (including subseries Lecture Notes in Artificial Intelligence and
  Lecture Notes in Bioinformatics)}, vol. 8486 LNCS, 2014, pp. 79--88.

\bibitem{Sem03}
\BIBentryALTinterwordspacing
C.~Semple and M.~Steel, \emph{Phylogenetics}, ser. Oxford lecture series in
  mathematics and its applications.\hskip 1em plus 0.5em minus 0.4em\relax
  Oxford University Press, 2003. [Online]. Available:
  \url{https://books.google.gr/books?id=uR8i2qetjSAC}
\BIBentrySTDinterwordspacing

\bibitem{Exo96}
\BIBentryALTinterwordspacing
G.~Exoo, ``\BIBforeignlanguage{eng}{A simple method for constructing small
  cubic graphs of girths 14, 15, and 16.}'' \emph{\BIBforeignlanguage{eng}{The
  Electronic Journal of Combinatorics}}, vol.~3, no.~1, p.~30, 1996. [Online].
  Available: \url{http://eudml.org/doc/119250}
\BIBentrySTDinterwordspacing

\bibitem{Har73}
\BIBentryALTinterwordspacing
J.~A. Hartigan, ``Minimum mutation fits to a given tree,'' \emph{Biometrics},
  vol.~29, no.~1, pp. 53--65, 1973. [Online]. Available:
  \url{http://www.jstor.org/stable/2529676}
\BIBentrySTDinterwordspacing

\bibitem{Fli90}
\BIBentryALTinterwordspacing
C.~Flight, ``How many stemmata?'' \emph{Manuscripta}, vol.~34, no.~2, pp.
  122--128, 1990. [Online]. Available:
  \url{http://dx.doi.org/10.1484/J.MSS.3.1335}
\BIBentrySTDinterwordspacing

\bibitem{Oei10}
\BIBentryALTinterwordspacing
N.~Sloane. (2010) {The On-Line Encyclopedia of Integer Sequences}. [Online].
  Available: \url{http://oeis.org}
\BIBentrySTDinterwordspacing

\bibitem{Bon98}
\BIBentryALTinterwordspacing
M.~Bonet, M.~Steel, T.~Warnow, and S.~Yooseph, ``{Better methods for solving
  parsimony and compatibility},'' \emph{J Comput Biol}, vol.~5, no.~3, pp.
  391--407, 1998. [Online]. Available:
  \url{http://www.ncbi.nlm.nih.gov/pubmed/9773340}
\BIBentrySTDinterwordspacing

\bibitem{Her04}
E.~a. Herniou, J.~a. Olszewski, D.~R. O'Reilly, and J.~S. Cory, ``{Ancient
  coevolution of baculoviruses and their insect hosts.}'' \emph{Journal of
  virology}, vol.~78, no.~7, pp. 3244--3251, 2004.

\bibitem{Pie02}
\BIBentryALTinterwordspacing
W.~H. Piel, M.~Donoghue, and M.~Sanderson, ``{TreeBASE : A database of
  phylogenetic information},'' in \emph{Proceedings of the 2nd International
  Workshop of Species 2000}, 2002, pp. 41--47. [Online]. Available:
  \url{http://phylogeny.harvard.edu/treebase.}
\BIBentrySTDinterwordspacing

\bibitem{Fel89}
J.~Felsenstein, ``{Phylip: phylogeny inference package (version 3.2)},''
  \emph{Cladistics}, vol.~5, pp. 164--166, 1989.

\bibitem{Swo03}
\BIBentryALTinterwordspacing
D.~L. Swofford, ``{Phylogenetic Analysis Using Parsimony},'' \emph{Options},
  vol.~42, pp. 294--307, 2003. [Online]. Available:
  \url{http://www.springerlink.com/index/10.1007/BF02198856}
\BIBentrySTDinterwordspacing

\bibitem{Yan03}
M.~Yan and D.~A. Bader, ``D.a.: Fast character optimization in parsimony
  phylogeny reconstruction,'' Georgia Institute of Technology, Tech. Rep.,
  2003.

\bibitem{Bad06}
D.~A. Bader, V.~P. Chandu, and M.~Yan, ``{ExactMP: An efficient parallel exact
  solver for phylogenetic tree reconstruction using maximum parsimony},'' in
  \emph{Proceedings of the International Conference on Parallel Processing},
  2006, pp. 65--73.

\bibitem{Sri08}
S.~Sridhar, F.~Lam, G.~E. Blelloch, R.~Ravi, and R.~Schwartz, ``{Mixed integer
  linear programming for maximum-parsimony phylogeny inference},'' in
  \emph{IEEE/ACM Transactions on Computational Biology and Bioinformatics},
  vol.~5, no.~3, 2008, pp. 323--331.

\bibitem{Goe08}
A.~Goeffon, J.~M. Richer, and J.~K. Hao, ``{Progressive tree neighborhood
  applied to the maximum parsimony problem},'' \emph{IEEE/ACM Transactions on
  Computational Biology and Bioinformatics}, vol.~5, no.~1, pp. 136--145, 2008.

\bibitem{Alo10}
N.~Alon, B.~Chor, F.~Pardi, and A.~Rapoport, ``{Approximate maximum parsimony
  and ancestral maximum likelihood},'' \emph{IEEE/ACM Transactions on
  Computational Biology and Bioinformatics}, vol.~7, no.~1, pp. 183--187, 2010.

\bibitem{Whi11}
W.~T.~J. White and B.~R. Holland, ``{Faster exact maximum parsimony search with
  XMP},'' \emph{Bioinformatics}, vol.~27, no.~10, pp. 1359--1367, 2011.

\end{thebibliography}

\fullversion{All links were last followed on February 10, 2016.
}

\end{document}